\documentclass[aps,pre,groupedaddress,notitlepage]{revtex4}
\usepackage{graphicx,amsmath,amssymb}
\usepackage[usenames,dvipsnames]{xcolor}

\begin{document}
\title{
Anomalous dynamics of intruders in a crowded environment of mobile obstacles
}
\author{Tatjana Sentjabrskaja$^{1,\S}$}
\author{Emanuela Zaccarelli$^{2,3,\S}$}
\email[]{emanuela.zaccarelli@cnr.it}
\author{Cristiano De Michele$^3$}  
\author{Francesco Sciortino$^{2,3}$}
\author{Piero Tartaglia$^3$}
\author{Thomas Voigtmann$^{4,5}$}
\author{Stefan U. Egelhaaf$^1$}
\author{Marco Laurati$^{1,6}$}
\email[]{marco.laurati@uni-duesseldorf.de}
\affiliation{$^1$ Condensed Matter Physics Laboratory, Heinrich Heine University, 40225 D{\"u}sseldorf, Germany}
\affiliation{$^2$ CNR-ISC, Universit\`{a} di Roma ``La Sapienza'', Piazzale A. Moro $2$, $00185$ Roma, Italy}
\affiliation{$^3$ Dipartimento di Fisica, Universit\`{a} di Roma ``La Sapienza'', Piazzale A. Moro $2$, $00185$ Roma, Italy}
\affiliation{$^4$ Institut f\"ur Materialphysik im Weltraum, Deutsches Zentrum f\"ur Luft- und Raumfahrt (DLR), 51170 K\"oln, Germany}
\affiliation{$^5$ Heinrich Heine University, Universit\"atsstra\ss{}e 1, 40225 D\"usseldorf, Germany}
\affiliation{$^6$ Divisi{\'o}n de Ciencias e Ingenier{\'i}a, Universidad de Guanajuato, Loma del Bosque 103, 37150 Le{\'o}n, Mexico}
\maketitle

{\bf
Many natural and industrial processes rely on constrained transport, such as proteins moving through cells,  particles confined in nanocomposite materials or gels, individuals in highly dense collectives and vehicular traffic conditions.  These are examples of motion through crowded environments, in which the host matrix may retain some glass-like dynamics.  Here  we investigate constrained transport in a colloidal model system, in which dilute small spheres move in a slowly rearranging, glassy matrix of large spheres. Using confocal differential dynamic microscopy and simulations, we discover a critical size asymmetry at which anomalous collective transport of the small particles appears, manifested as a logarithmic decay of the density autocorrelation functions.  We demonstrate that the matrix mobility is central for the observed anomalous behaviour.  These results, crucially depending on size-induced dynamic asymmetry, are of relevance for a wide range of phenomena ranging from glassy systems to cell biology.
\\
}

In the presence of a confining medium, the transport of objects deviates from normal diffusion.
Anomalous behaviour, usually manifested by the presence of sub-diffusivity~\cite{klafter,Hofling_PhysRep}, emerges as a common feature of the dynamics.   In the Lorentz gas \cite{Lorentz,franosch_frey}, 
the prototype model for anomalous transport, point-like intruders move in voids between immobile, randomly-distributed particles. Their motion becomes sub-diffusive once the voids are barely interconnected. 
When a critical density of immobile particles  is reached, they percolate and the intruder becomes localized \cite{Lorentz}.
Softness of the immobile particles or interactions among the intruders are known to modify this picture~\cite{krackoviak,coslovich,saito,voigtmann_horbach09,schnyder15,horbach_dullens}.

\begin{figure}[tb]
\centerline{\includegraphics[scale=0.4]{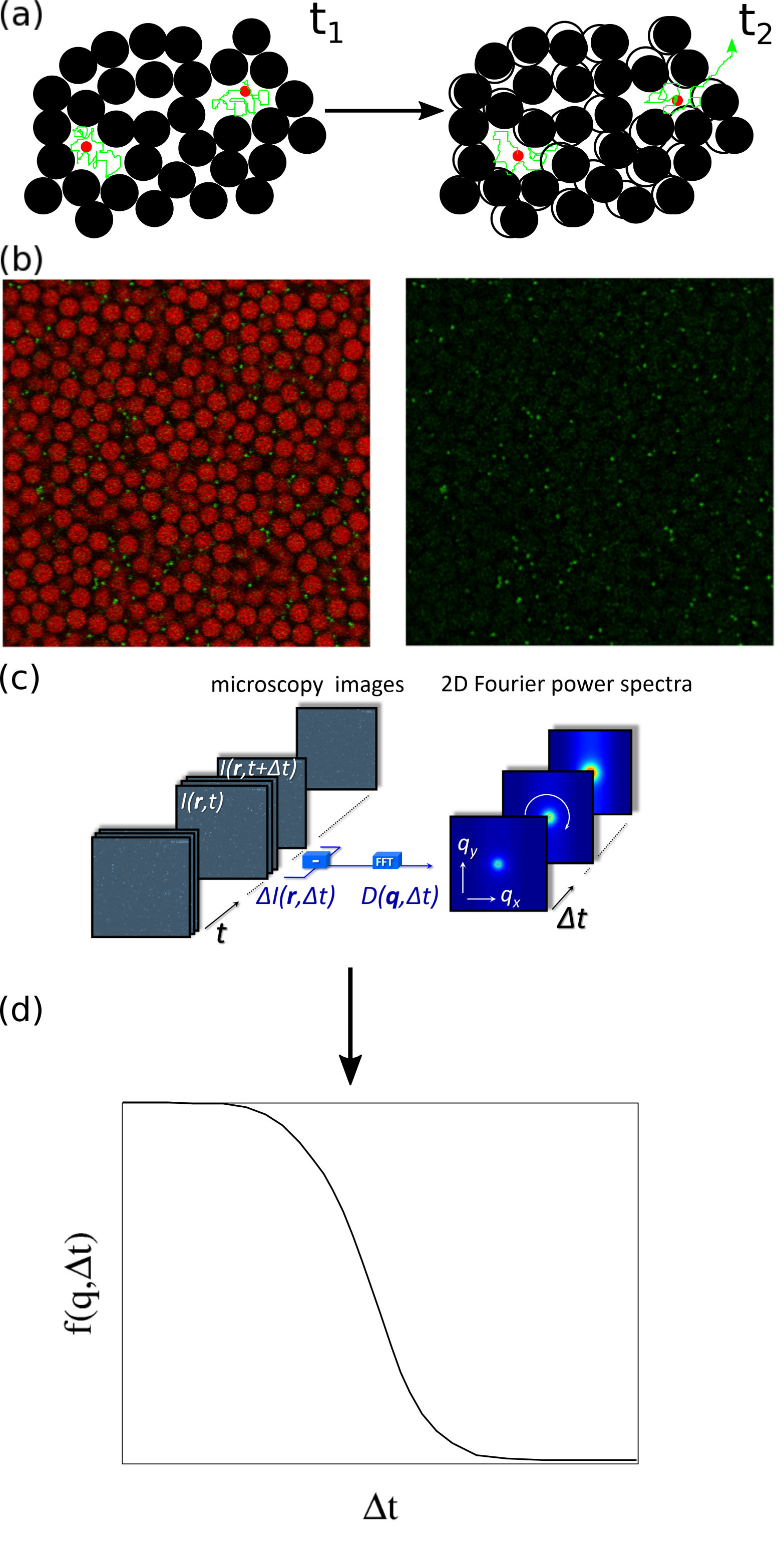}}
\caption{\label{fig1}
{\bf Illustration of the system and measurement method} (a) Schematic illustration of our system at two times $t_1$ and $t_2>t_1$ highlighting the trajectories (green lines) of the intruders (red beads) in voids and between voids made possible due to the mobility of the matrix particles. (b) An exemplary confocal microscopy image of a mixture with $\delta$ = 0.18 and $\phi$ = 0.58 in which (left) both particles and (right) only the small particles are shown. (c) Image differences at different delay times $\Delta t$ are Fourier transformed to give 2D Fourier power spectra for different $\Delta t$. (d) After azimuthal averaging and additional treatment the intermediate scattering function $f(q, \Delta t)$ is obtained.
}
\end{figure}

So far the slow movement of the host matrix has been largely ignored, despite representing
realistic situations of biological \cite{Ellis_Nature,dirienzo_ncomm,sadati,angelini,trimble,gravish2015} and industrial interest \cite{cherdhirankorn2009,grabowski2014,narayanan_2007,kalathi_2014,babu_2008,salami_2013,helbing2001}. To address confined transport in slowly moving matrices, here we investigate a binary colloidal mixture  of small and large hard spheres, of diameters $\sigma_\mathrm{s}$ and $\sigma_\mathrm{l}$, which represent  intruders and  host matrix, respectively. 
Changing the size ratio $\delta=\sigma_\mathrm{s}/\sigma_\mathrm{l}$ we also modify the dynamic asymmetry of the system. We focus on volume fractions of large particles $\phi_l>0.5$ approaching the glass transition, occurring at $\phi_l^g\approx 0.58$. In contrast the volume fraction of the intruders $\phi_\mathrm{s}$ is very small with $x_\mathrm{s} \equiv \phi_\mathrm{s}/\phi = 0.01 $. 
Such a system combines the confinement of a dilute fluid of mobile intruders 
with the slow dynamics of the matrix (Fig.~\ref{fig1}a). It thus provides the simplest minimal model for 
the  investigation of motion in crowded soft and biological matter.

Despite its conceptual simplicity, experimental investigations of the dynamics of small intruders in mixtures of  Brownian particles with large size-asymmetry are scarce.
This might be due to limitations in the spatial and temporal resolution of confocal microscopy which make it difficult to track particles that are significantly smaller than another species of Brownian, i.e. at most micron-sized, particles. 
To overcome these limitations, we keep the selectivity of fluorescent labelling (Fig.\ref{fig1}b), which allows us to separately determine the small and large particles. However, instead of tracking we employ the recent Differential Dynamic Microscopy (DDM) technique \cite{cerbino_prl,lu,poonDDM}. This is based on the time correlation in Fourier space of the difference between images separated by a time delay $\Delta t$ (Fig.\ref{fig1}c) and provides a measure of the (isotropic) collective intermediate scattering function or density autocorrelation function $f(q,\Delta t)$, where $q$ is the modulus of the wavevector ${\bf q}$ (Fig.\ref{fig1}d). The decay of $f(q,\Delta t)$ as a function of time delay $\Delta t$ corresponds to the loss of correlation of the particle density on a length scale determined by $q^{-1}$ within the time delay $\Delta t$. The decay time is therefore related to the characteristic time of the particle motions on the length scale $q^{-1}$. Approaches similar to DDM, like fluorescence correlation spectroscopy, do not provide information on the probed length scale. This information is crucial to investigate the effect on the dynamics of the size of the voids in which the small particles move. The function $f(q,\Delta t)$ can also be obtained by dynamic light scattering, which, however, does not allow us to distinguish the two species by fluorescent labeling. We also study the same system by mode coupling theory of the glass transition (MCT) and, both in the case of mobile and immobile matrix particles, by numerical simulations, complementing the experimental results and providing insights on the underlying microscopic mechanisms. We observe anomalous dynamics of the small spheres at a critical size ratio $\delta_\mathrm{c}$ and we show that this dynamical behavior is intimately connected to the slow dynamics of the matrix of large particles. \\

\begin{figure}[tb]
\centerline{\includegraphics[scale=0.38]{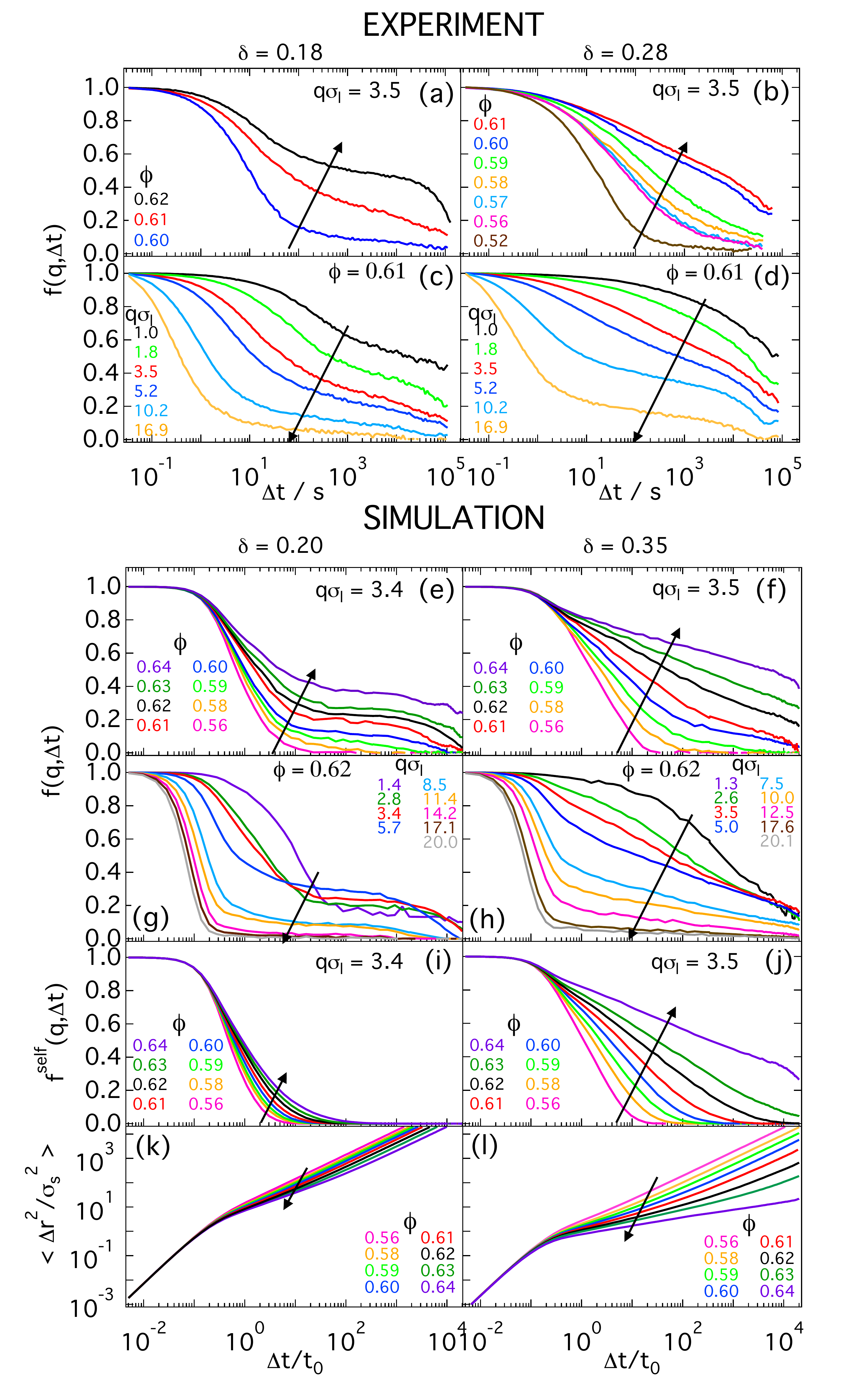}}
\caption{\label{fig2}
{\bf Dynamics of the intruders as observed in experiments and simulations}
Collective $f(q,\Delta t)$ (a--h) and self $f^{self}(q,\Delta t)$ (i--j) intermediate scattering functions and mean-squared displacements $\langle\Delta r^2/\sigma_\mathrm{s}^2\rangle$ (k--l) as a function of delay time $\Delta t$, describing the dynamics of small spheres in binary mixtures with size ratios $\delta$ below (left) and around (right) the onset of anomalous dynamics, for different magnitudes of the scattering vector $q$ and total volume fraction $\phi$ (as indicated).
}
\end{figure} 

{\bf Results\\}

{\bf Small particle dynamics\\}

Fig.~\ref{fig2}a-d shows 
the measured collective intermediate scattering functions $f(q,\Delta t)$ of the small particles  for size ratios $\delta=0.18$ (Fig.\ref{fig2}a,c)  and $\delta=0.28$ (Fig.\ref{fig2} b,d)  for different $\phi$ and $q$.
For $\delta =  0.18$ and 
all $\phi$ and $q$, $f(q,\Delta t)$ vs. $\Delta t$  shows an initial decay, followed by a $\phi$-dependent intermediate plateau, and eventually a decay to zero at longer times (Fig.\ref{fig2}a). The initial decay can be associated with the Brownian motion of small particles within the voids of the large particles matrix. It becomes increasingly slower for increasing $\phi$ (Fig.\ref{fig2}a) and decreasing $q$, which means increasing length scale (Fig.\ref{fig2}c). 
The intermediate plateau indicates the dynamical arrest of the collective dynamics, i.e. of density fluctuations, and hence the absence of diffusion on the length scale determined by $q^{-1}$. 
 The height of the plateau increases progressively with increasing $\phi$, similarly to the scenario in which a
percolation-type transition is approached~\cite{MCTAType,krackoviak}, and indicates that voids become smaller and particles are increasingly localised~\cite{BosseOLDstuff}. The final decay to zero of $f(q,\Delta t)$ shows that particles are still able to diffuse at long times. For a larger size ratio, $\delta =  0.28$, and comparable $\phi$ values, a completely different scenario appears. Beyond $\phi \approx 0.60$, $f(q,\Delta t)$ shows remarkable anomalous dynamics, manifested in an extended logarithmic decay over three decades in time. This intriguing behavior is mostly visible at $\phi\gtrsim 0.60$ and $q\sigma_\mathrm{l} \approx 3.5$, i.e. when probing a length scale of about $2\sigma_\mathrm{l}$ (Fig.\ref{fig2}b), which is comparable to the size of the matrix particles.



The experimental findings are confirmed by simulations. For $\delta=0.20$ no anomalous behavior of the small particles is detected in the collective $f(q,\Delta t)$  (Fig.\ref{fig2}e,g) and in the self  $f^\mathrm{self}(q,\Delta t)$ correlation functions (Fig.\ref{fig2}i). 
Note that for $\delta = 0.20$, $f(q,\Delta t)$   displays a two step-relaxation and the presence of localisation (Fig.\ref{fig2}e,g), which is absent in $f^\mathrm{self}(q,\Delta t)$ (Fig.\ref{fig2}i). Also the mean squared displacements (MSD) $\langle \Delta r^2\rangle\equiv \langle |\vec r(t)- \vec r(0)|^2\rangle$, with $\vec r(t)$ the position of a particle at time $t$, show almost no localisation at all $\phi$ (Fig.\ref{fig2}k). 
This decoupling between collective ($f(q,\Delta t)$) and self dynamics ($f^\mathrm{self}(q,\Delta t)$, MSD) originates from the glassy environment in which the intruders move. Correlated motions of a group of intruders distributed within the matrix are more influenced by the slow dynamics of the matrix particles than uncorrelated single particle motions, which are mostly sensitive to the local structure of the voids~\cite{krackoviak,Thomas_EPL}.
For $\delta=0.35$ we find the emergence of logarithmic anomalous relaxations of $f(q,\Delta t)$ (Fig.~\ref{fig2}f,h) and $f^\mathrm{self}(q,\Delta t)$ (Fig.\ref{fig2}j), for comparable $q$ as in the experiments. Additional simulations for $\delta=0.30$ and $\delta=0.40$ also show a logarithmic decay over a smaller time window. Furthermore, for $\delta=0.35$ and $\phi \gtrsim 0.60$ the MSD displays a clear sub-diffusive behavior, i.e. $ \langle \Delta r^2\rangle \sim t^{\alpha}$ with $\alpha <1$ (Fig.~\ref{fig2}l). Finally, for $\delta=0.5$, $f(q,\Delta t)$ and $f^\mathrm{self}(q,\Delta t)$ show a two-step decay and the MSD a localisation plateau at large $\phi$, consistent with a standard glass transition of the small particles. At all investigated $\delta$ and for  $\phi_l>0.55$, the dynamics of the large particles  are very slow and at intermediate times are indicating localisation and motion within nearest neighbour cages of approximate size $0.1 \sigma_\mathrm{l}$ (Supplementary Fig. 1). 

\begin{figure}[tb]
\centering
\includegraphics[scale = 0.6]{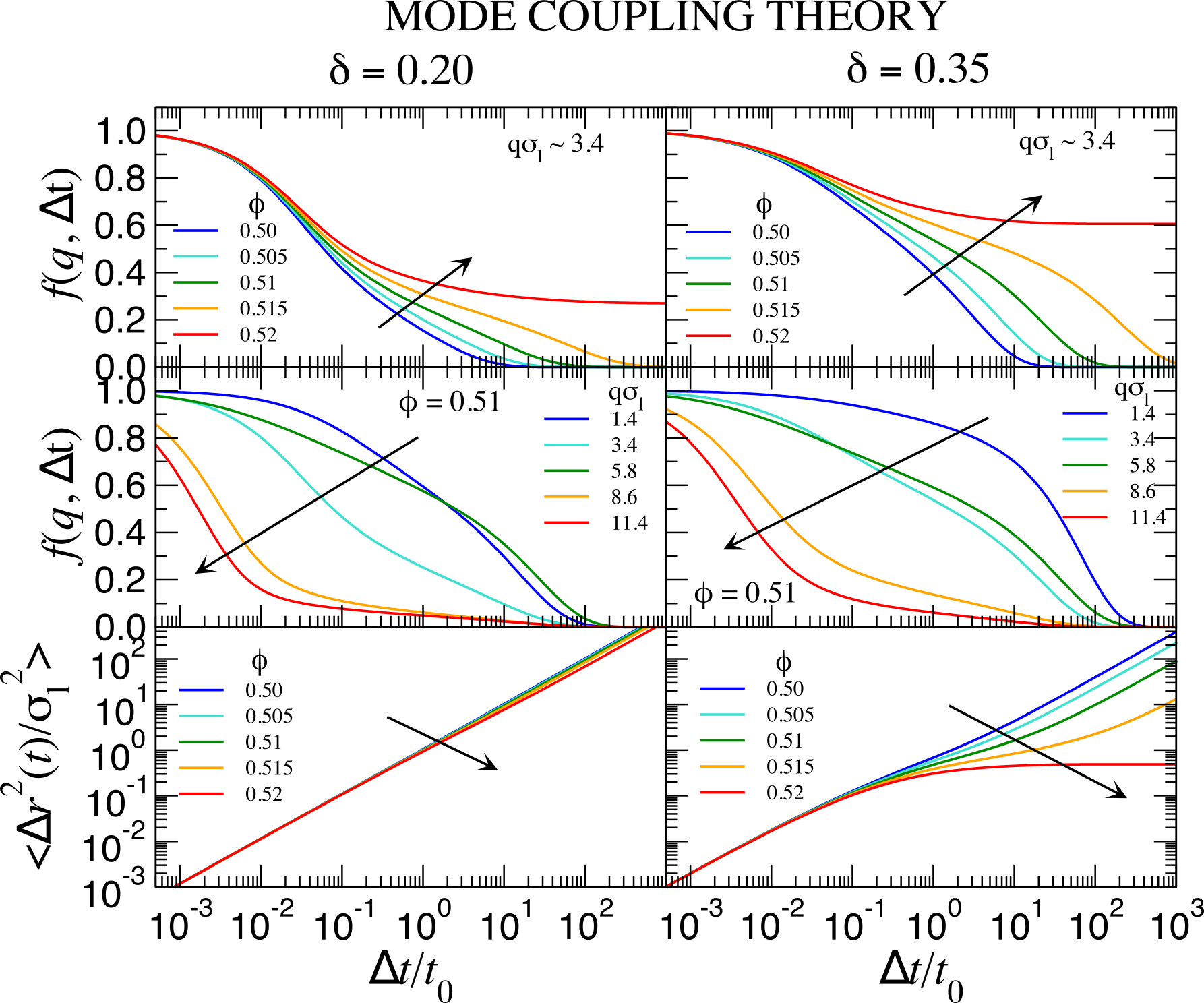}
\caption{
  {\bf Dynamics of the intruders as predicted by MCT} Intermediate scattering functions
$f(q, \Delta t)$ (top, middle) and mean-squared displacements $\langle \Delta r^2/\sigma_\mathrm{l}^2\rangle$ (bottom) describing the dynamics of small spheres in binary mixtures with size ratios delta below (left) and around (right) the onset of anomalous dynamics, for different magnitudes of the scattering vector $q$ and total volume fraction $\phi$ (as indicated).
\label{fig:mct}
}
\end{figure} 

These results suggest the existence of a critical size ratio $\delta_c\simeq 0.35$ at which pronounced   
anomalous dynamics mark the transition from a diffusive to a glassy regime of the small particles moving in the large particles matrix. 
The $\delta_c$ and $\phi$ values where this transition is observed are slightly smaller in the experiments than in the simulations. This is attributed to the fact that in the experiments small particles are polydisperse, while in the
simulations they are monodisperse. Polydispersity is expected to affect the transition since the average size particles might still be able to diffuse through the void spaces in the matrix, whereas the largest particles of the size distribution might no longer be able to diffuse through them.
The crossover observed at $\delta_c$ is analogous to the transition from a diffusive to a localized state in models with fixed obstacles. 
However, the excluded volume of the intruder generates a coupling with the host matrix and, due to the mobility of the matrix, also between intruders in different voids, mutating localization into a glass transition due to the (slow) mobility of the matrix particles. Although this is apparently similar to 
intruders in a fixed matrix~\cite{krackoviak,coslovich,saito}, the logarithmic decay of $f(q,\Delta t)$ stands out as a novel feature.

On the basis of mode coupling theory (MCT), the appearance of logarithmic decays in $f(q,\Delta t)$ \cite{Moreno_jcp,Moreno_jcp2,MayerMacromolecules} is usually attributed to competing collective arrest mechanisms, like caging and bonding, and to higher-order glass transition singularities \cite{Dawson00,Sciortino03,Gnan14,Thomas_EPL}. We solved MCT equations for a binary mixture of hard spheres and $x_\mathrm{s} = 0.01$. The resulting correlators $f(q,\Delta t)$ for a range of
packing fractions around the MCT glass transition, $\phi_\mathrm{c}\approx 0.516$ and $\delta = 0.20$ and 0.35,
are shown in Fig. \ref{fig:mct}. No clear sign of logarithmic decay of $f(q,\Delta t)$ is found
for these states in MCT: while an approximate logarithmic dependence of the decay  is observed at $\delta = 0.35$, $\phi = 0.51$ and $q\sigma_\mathrm{l} = 3.4$, this extends over an interval of times much shorter than in experiments and simulations. In addition, upon further increasing $\phi$ the logarithmic dependence does not take over, but instead a two-step decay is found, followed by the arrest of the dynamics. Indeed higher-order singularities are not present in this region of $\phi$ and $x_\mathrm{s}$ values \cite{Thomas_EPL}. On the other hand, the MSD obtained from MCT shows the qualitative signatures found in simulations:
for $\delta=0.20<\delta_\mathrm{c}$, the long-time diffusion barely slows down with
increasing $\phi$, indicating a partially frozen glass in which the small particles are mobile. For $\delta=0.35\approx\delta_\mathrm{c}$, anomalous sub-diffusion is observed, indicating that the glass-transition of the large particles and the localization transition of the small particles are close to each other. Thus, the appearance of approximately logarithmic decay in Fig. \ref{fig:mct} could be a signal of the transition from coupled dynamics of the two species at large $\delta$ to decoupled dynamics at small $\delta$.\\ 

\begin{figure}[tb]
\centerline{\includegraphics[scale=0.28]{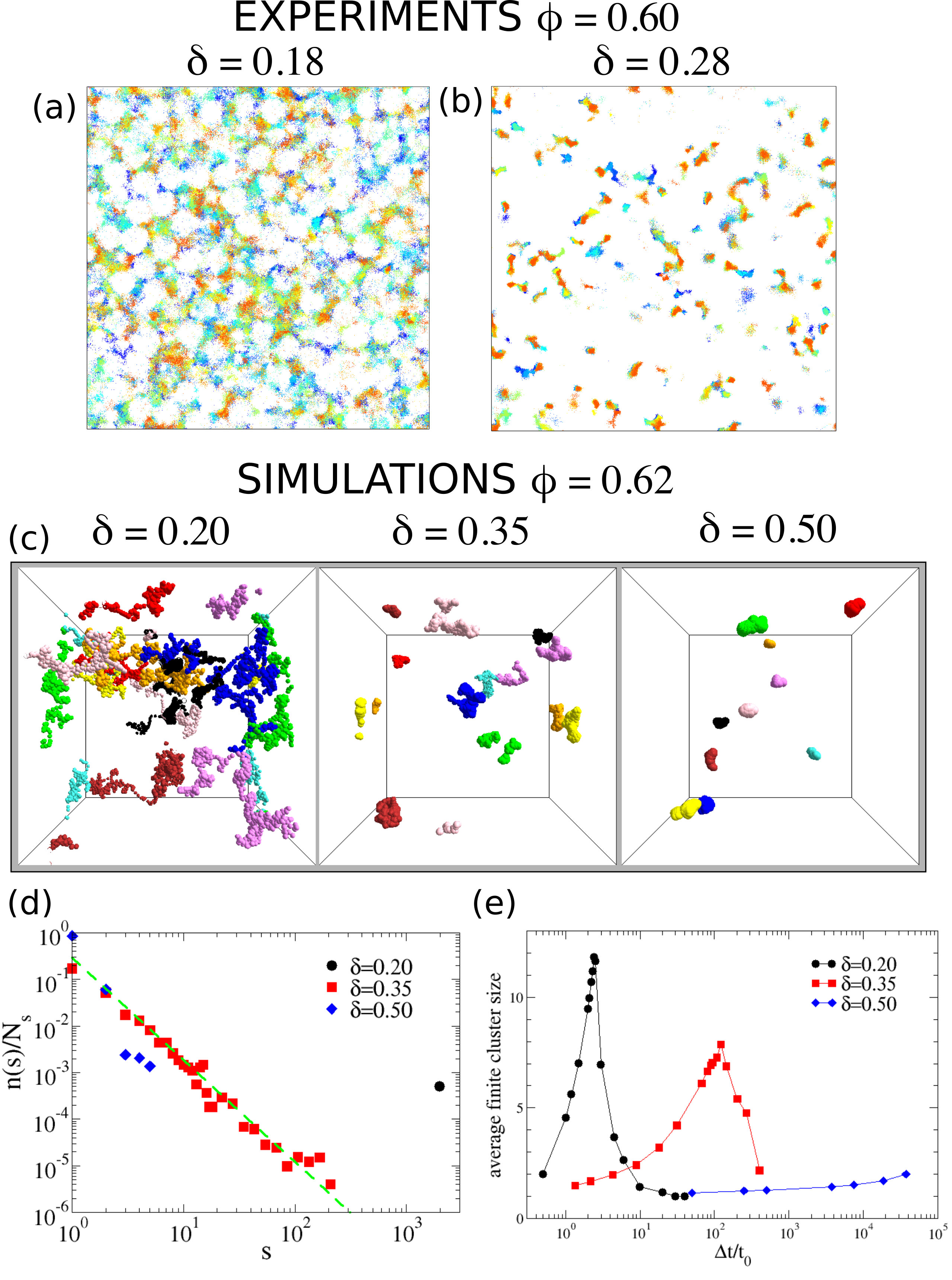}}
\caption{\label{fig4} 
{\bf Illustration of the space explored by small particles during their motions}
(a-b) Overlay of small particle positions at different times (colour coded from blue, corresponding to $t_i^\mathrm{exp} = 0$~s to red, corresponding to $t_f ^\mathrm{exp}=297$~s with time steps of $33$~s), obtained by particle tracking applied to 2D confocal microscopy images, for $\phi = 0.60$ and (a) $\delta = 0.18$, (b) $\delta = 0.28$. 
(c) Positions of ten small particles (distinguished by different colours) for (left) $\delta=0.2$, (middle) $\delta=0.35$,  and (right) $\delta=0.5$, for a fixed total time of the trajectories $t_f^\mathrm{sim}=100t_0$, comparable to the experiments
r
(d) Distribution $n(s)$ (normalized by the number of small particles $N_s$) of the size $s$ of the space explored by small particles, evaluated within a fixed time interval $t_f^\mathrm{sim}=100t_0$. For $\delta=0.35$ data are consistent with a power-law dependence $n(s)\sim s^{-2.19}$, consistent with random percolation (dashed line), while for $\delta=0.20$ all particles belong to the same cluster. 
(e) Average size $L_\mathrm{c}$ of finite clusters as a function of time, for different $\delta$, as indicated. The maximum in each curve signals the onset of percolation.}
\end{figure} 

{\bf Void space explored by small particles\\} 

A direct visualisation of small particle locations shows that the transition from diffusive dynamics at small $\delta$ to localised dynamics at large $\delta$ observed in experiments, simulations and theory is associated, similarly to models with immobile obstacles, with the transition from percolating to non-percolating voids within the matrix. However, 
a static picture of the void geometry cannot describe this transition, because the evolution of the void space involves a second timescale $t_2$ (Fig.~\ref{fig1}a, right) associated with the mobility of the matrix.
To analyse the dynamic rearrangements of the void structure, we monitor the evolution of the position of the small particles which explore this evolving structure.
Accordingly, in Fig.~\ref{fig4}a,b we show superpositions of small particle locations in 2D time series of confocal images over a long total observation time $t_f^\mathrm{exp}= 297$~s, at which $f(q,\Delta t)$ for $\delta = 0.18$ shows a decay of correlations, while $f(q,\Delta t)$ for $\delta = 0.28$ is in the logarithmic regime.
For $\delta = 0.18$ we find that, within the observation time, small particles easily explore the whole space of the accessible voids which form a percolating network. In contrast, for $\delta = 0.28$ particles mostly explore their local environment, since voids only barely connect even at long times, allowing only a slow, partial exploration of the available void space.
Simulations provide not only particle locations but also single-particle trajectories in three dimensions allowing a more quantitative determination of the percolation of the explored space. Visualisations of typical small particle trajectories for a fixed observation time $t_f^\mathrm{sim} = 100 t_0$ (comparable to the experiments) and three different values of $\delta$ confirm the experimental features (Fig.\ref{fig4}c): within the observation time small particles explore a percolated space  for small $\delta$, while for the critical size ratio the space is barely connected, indicating that particles can rarely escape the local environment which is only possible due to the stochastic opening and closing of channels between neighbouring void spaces, associated with the matrix motion on the long time scale t2. In addition the simulations show that for even larger $\delta$ the explored space is disconnected.  
To quantify these observations we calculate the size distribution $n(s)$ of the space $s$ explored by small particles within a certain time interval, as explained in Methods. The results are shown in Fig.\ref{fig4}d for different $\delta$ values for an observation time equal to $t_f^\mathrm{sim}$. This time corresponds to the interval over which the cluster size distribution of the explored space for $\delta_c$ is close to percolation, as indicated  by the power-law dependence $n(s)\sim s^{-2.19}$, consistent with random percolation predictions  \cite{stauffer}. 
Percolation at $t_f^\mathrm{sim}$ for $\delta_\mathrm{c}$ is also indicated, in a finite-size system, by the maximum of the average size of finite-size clusters (excluding percolating clusters, calculated as explained in Methods) $L_\mathrm{c}$ as a function of time (Fig.\ref{fig4}e). For the other size ratios instead $L_\mathrm{c}$ is very small at $t_f^\mathrm{sim}$.  
At small $\delta$ this is due to the fact that particles can easily move through channels connecting voids, and thus the explored space quickly associates into a percolating cluster.
On the other hand, for large $\delta$ the creation of channels that allow the small particles to move between neighbouring void spaces is rare, and thus percolation of the explored space does not occur at $t_f^\mathrm{sim}$ and only voids corresponding to the size of monomers, dimers and few-mers are observed.
This analysis reveals very different timescales at which the explored space percolates at different $\delta$. These timescales depend, besides $\delta$, on the timescale $t_2$ of the evolution of the void space, associated with the thermal motion of the matrix particles: yet this analysis is not offering
substantial evidence that this mobility of the matrix is causing the logarithmic decays of the correlators observed at $\delta_\mathrm{c}$.\\

 \begin{figure}[tb]
\centerline{\includegraphics[scale=0.4]{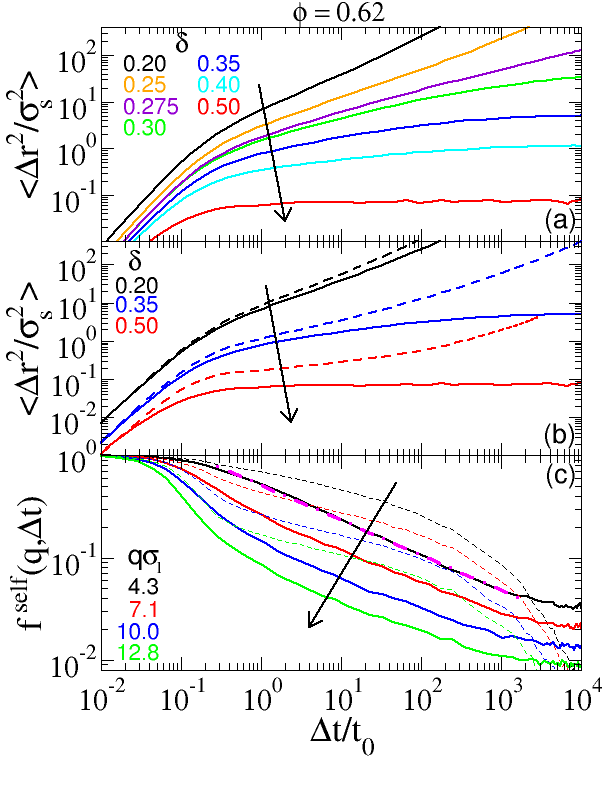}}
\caption{\label{immobileMSD}
{\bf Small particle dynamics in a mobile or immobile large particles matrix}
(a) MSDs of the small particles for immobile large particles at $\phi=0.62$ and various values of $\delta$, as indicated. 
For $\delta_c^\mathrm{imm}\sim 0.275$ a clear subdiffusive behavior is observed at all times.
(b) Comparison of the MSDs of the small particles at $\phi=0.62$ for mobile (dashed lines) and immobile (full lines) large particles, for increasing $\delta$, as indicated. 
(c) Self intermediate scattering functions $f^\mathrm{self}(q,\Delta t)$ at $\phi=0.62$ and different wavevectors $q\sigma_\mathrm{l}$, as indicated, for $\delta=0.25$ (immobile, full lines) and $\delta=0.35$ (mobile, dashed lines) highlighting the power-law dependence (dot-dashed line) in the immobile case.}
\end{figure}

{\bf Comparison between mobile and immobile matrix\\} 

To go one step further 
and link the residual mobility of the matrix particles with the anomalous logarithmic decays, we perform additional simulations (for $\phi=0.62$) 
for immobile matrix particles and compare the dynamics of the intruders with the case of a mobile matrix. 
When the large particles are immobile  (Fig.~\ref{immobileMSD}a), the MSD shows a sub-diffusive regime 
(MSD$\sim t^\alpha$) followed by diffusion at long times (upward curvature) or localization
(downward curvature), depending on $\delta$. The crossover between these two long time behaviors  takes place at a critical size ratio
$ \delta_\mathrm{c}^\mathrm{imm} \sim 0.275$  where the MSD remains subdiffusive also at long times \cite{Hofling_PhysRep}.
The value of $\delta_c$ is smaller for  the simulation with immobile large particles. This finding is 
consistent with the opening of channels as a consequence of the thermal motion of the matrix particles. 
In the case of mobile matrix particles 
localisation is {\it never} observed (Fig.~\ref{immobileMSD}b): 
even for large $\delta$, the residual motion
of the matrix allows the small particles to move and hence their MSD increases at long times. Furthermore, the subdiffusive regime is only observed for $\delta < \delta_\mathrm{c}^\mathrm{imm}$ and thus in a smaller range than for mobile particles. This is consistent with the opening of channels as a consequence of the thermal motion of the matrix particles, which allows larger particles to move between voids.
We also find that   $f^\mathrm{self}(q,\Delta t)$ calculated for the case of an immobile matrix displays a power-law dependence on time  extending for several decades (Fig.~\ref{immobileMSD}c), as also observed in the Lorentz gas model~\cite{franosch_softmatter},  while the collective $f(q,\Delta t)$ displays neither a power-law nor a logarithmic dependence (Supplementary Fig.2).  In the case of a mobile matrix, however, power law behaviour is not observed but, close to $\delta_\mathrm{c}$,  a logarithmic dependence is found. Thus, thermal motion of the matrix  particles gives rise to the logarithmic decay,  a novel type of dynamics which does not occur in models with immobile obstacles.\\

{\bf Discussion\\}

Our combined experimental, simulation and theoretical study shows that dynamics of 
intruders in a mobile crowded environment 
requires a description beyond that provided by models with a matrix of fixed obstacles. 
The novel application of the confocal DDM technique to concentrated binary colloidal mixtures  allows us to investigate the collective dynamics of intruders  in a mobile matrix, revealing extended anomalous dynamics for specific
values of the size  asymmetry and of the probed length scale. While the Lorentz model predicts a
 power-law behavior, which is typical for systems close to a percolation transition,
 in the case of a mobile matrix we observe a logarithmic decay of the collective and self density fluctuations over  at least three decades in time,
 at length scales comparable to the
size of the matrix particles. 
This logarithmic decay marks the transition between  a diffusive behaviour of intruders in a glassy medium 
for small size ratios $\delta < \delta_\mathrm{c}$, where
transient localization is due to the excluded volume of the mobile matrix, and glassy dynamics of the intruders at large
size ratios $\delta > \delta_\mathrm{c}$, due to crowding.  
 Our results thus show that both percolation and glassy dynamics have to be considered.
By comparing mobile and immobile matrix environments, we demonstrate  that the dynamics of the small particles is profoundly altered, in a qualitative way, by the continuous evolution of channels in the mobile matrix, due to the thermal motion of large particles. 
A mobile matrix corresponds to an environment in which small intruders move in many real systems and applications, like in glasses, nanocomposite materials, chromatography, catalysis, oil recovery, drug delivery or cell signaling, cell interiors, human and animal crowds and vehicular traffic. We thus expect that our findings will inspire the development of a more realistic description of these situations and will stimulate theoretical studies to refine the MCT predictions. 

\section*{Methods}

{\bf Materials} We investigated dispersions of sterically
stabilized PMMA spheres of diameters $\sigma_{\mathrm{l}^{(1)}}= 3.10$~$\mu$m (polydispersity 0.07) or $\sigma_{\mathrm{l}^{(2)}} = 1.98$~$\mu$m (polydispersity 0.07)
mixed with spheres of diameter $\sigma_{\mathrm{s}} = 0.56$~$\mu$m (polydispersity 0.13) (fluorescently labeled with nitrobenzoxadiazole (NBD)),
in a cis-decalin/cycloheptyl-bromide  mixture which
closely matches their density and refractive index. The size ratio of the mixtures is $\delta = 0.18$ ($\sigma_{\mathrm{l}^{(1)}}$) and  $\delta = 0.28$ ($\sigma_{\mathrm{l}^{(2)}}$), respectively. 
After adding salt (tetrabutylammoniumchloride), this system presents hard-sphere like
interactions \cite{yethiraj03,Poon/Weeks/Royall}. 
A sediment of the large spheres with $\phi = 0.65$ or of the small spheres with $\phi = 0.67$, as estimated from comparison with numerical simulations and experiments \cite{silescu, weeks}, is diluted to obtain one-component dispersions with desired volume fraction $\phi$.
Following a recent study \cite{poon2012}, the uncertainty $\Delta\phi$ can be as large or above 3$~$\%. 
Using the nominal volume fraction $\phi$ of the large spheres as a reference, the volume fraction of the samples containing the small particles are adjusted in order to obtain comparable linear viscoelastic moduli in units of the energy density $3k_{\mathrm{B}}T/4\pi R^3$, where $k_{\mathrm{B}}$ is the Boltzmann constant, T the temperature and R the particles' radius, while multiplying the frequency by the free-diffusion Brownian time $\tau_0 = 6 \pi \eta R^3/k_{\mathrm{B}}T$, where $\eta=$ 2.2~mPa$\,$s is the solvent viscosity.
In this way we obtain samples with comparable rheological properties and, according to the generalised Stokes-Einstein relation \cite{mason2000}, also dynamics and hence a similar location with respect to the glass transition. The comparable dynamics but different polydispersities of the one-component samples imply slightly different $\phi$.
Samples with different total volume fractions and a fixed composition, namely a fraction of small particles $x_{\mathrm{s}} = \phi_{\mathrm{s}} /\phi = 0.01$, where $\phi_{\mathrm{s}}$ is the volume fraction of small particles, are prepared by mixing the one-component samples.\\ 
 
{\bf DDM measurements} Confocal microscopy images were acquired in a
plane at a depth of approximately 30~$\mu$m from the coverslip.  Images with 512$\times$512 pixels,
corresponding to 107~$\mu$m $\times$ 107~$\mu$m, were acquired at a fast rate of 30 frames per second to follow the short-time dynamics and at a slow rate,
 between 0.07 and 0.33  frames per second, depending on sample, to follow the long-time dynamics. Image series were acquired using a Nikon A1R-MP confocal scanning unit mounted on a Nikon Ti-U inverted microscope, with a 60x Nikon Plan Apo oil immersion objective (NA = 1.40). The pixel size at this magnification is 0.21~$\mu$m $\times$ 0.21~$\mu$m. The confocal images were acquired with the maximum pinhole size allowed by the microscope, corresponding to a pinhole diameter of 255~$\mu$m.   
Time series of 10$^4$ images were acquired for 2 to 5 different volumes, depending on sample.\\

{\bf DDM analysis} Particle movements induce fluctuations of the fluorescence intensity in the images, $i(x,y,t)$, with $x$, $y$ the coordinates of a pixel in the image and $t$ the time at which the image was recorded. 
To obtain additional information on the characteristic length scales of particle motions, $i(x,y,t)$ can be Fourier transformed, yielding $\hat{i}({\bf q},t)$, with ${\bf q}$ the wave vector in Fourier space, and then differences of the Fourier transformed image intensities can be correlated (Fig. \ref{fig1}c) to obtain the image structure function $D({\bf q},\Delta t)$:
\begin{equation}
D({\bf q},\Delta t) = \langle |\hat{i}({\bf q},t+\Delta t)-\hat{i}({\bf q},t)|^2\rangle
\label{eq_diqt_dfm}
\end{equation}
where $\langle \rangle$ represents an ensemble average. This analysis technique is named Differential Dynamic Microscopy (DDM) \cite{cerbino_prl}. The intermediate scattering function $f({\bf q},\Delta t)$ (Fig. \ref{fig1}d) can be extracted from the image structure function:
\begin{equation}
D({\bf q},\Delta t) = A({\bf q})[1-f({\bf q},\Delta t)]+B({\bf q})
\label{eq_diqt_fqt}
\end{equation}
with $A(q)= N|\hat{K}(q)|^2S(q)$, where $N$ is the number of particles in the observed volume, $\hat{K}(q)$ is the Fourier transform of the Point-Spread Function of the microscope, $S(q)$ is the static structure factor of the system, and $B(q)$ accounts for the camera noise. The inverse of the wave vector $q$ determines the length scale over which the particle dynamics are probed. Thus  $f({\bf q},\Delta t)$ is obtained, similarly to dynamic light scattering (DLS)\cite{berne-pecora}, but  for the present system the advantage of DDM over DLS is that fluctuations of the incoherent fluorescence signal can be correlated, a possibility which is excluded by the requirement of coherence of light in DLS. Furthermore, use of a confocal microscope drastically reduces the amount of background fluorescence of the measurements, significantly improving the determination of $f({\bf q},\Delta t)$. The effect of particles moving in and out of the observation plane on $f({\bf q},\Delta t)$ was found to be negligible for all samples, as determined by the q-dependence of the relaxation times of the initial decay of $f({\bf q},\Delta t)$, where no plateau at small q values was observed \cite{lu,cerbino_dfm}.\\

{\bf Particle Localization}
Coordinates of the small particles were extracted from time series of 2-dimensional images using standard particle localization routines based on the centroiding technique \cite{crocker}. Only the particle positions at each time could be determined, not the full trajectories. Indeed the displacement of small particles during the time delay $\Delta t$ between two successive frames is comparable or larger than their diameter, which implies that identifying particles after a $\Delta t$ becomes too uncertain.\\

{\bf Simulations}
We perform event-driven Molecular Dynamics simulations\cite{DeMichele} in the $NVT$ ensemble in a cubic box with periodic
 boundary conditions for binary mixtures of hard spheres, of which the large components are $7\%$ polydisperse by a discrete Gaussian distribution\cite{Zacca09} and the small ones are monodisperse. For each studied $\delta$ we vary the total number of particles in the range of a few thousands. The number of small particles thus varies from $1980$ for $\delta=0.2$ to $292$ for $\delta=0.5$.
Mass and  length are measured in units of particle mass $m$, average large particle diameter $\sigma_\mathrm{l}$, whereas time is in units of $t_0=
 \sqrt{m \sigma_l^2/\kappa_B T}$, where $k_B$ is the Boltzmann constant and $T$ the temperature. 
 For the simulations with immobile hard spheres, after equilibration of the mixture, we freeze the large particles only. To roughly estimate the critical size ratio which demarcates the transition between diffusive and localized states, we averaged results over ten different matrix realizations. \\

{\bf Mode Coupling Theory}
The equations determining $f(q,t)$ and $\langle\Delta r^2(t)\rangle$
within MCT were solved
for a binary mixture of hard spheres within the Percus-Yevick approximation for
the static structure; for details on the theory and the numerical procedure, see Ref.~\cite{Thomas_EPL}. The $f(q,\Delta t)$ were obtained using a wave-number grid of equidistant steps $\Delta q=0.4/\sigma_{\mathrm l}$, with large-$q$ cutoff $q\sigma_{\mathrm l}=400$. Brownian dynamics is assumed with the short-time diffusion coefficients following the Stokes-Einstein relation; the diffusion coefficient of the large particles sets the unit of time $\tau_0$. In the calculations, the total packing fraction $\phi$ is varied, keeping $x_\mathrm{s}=\phi_s/\phi=0.01$ fixed.\\

{\bf Calculation of the size distribution of the explored space}
To evaluate the distribution of space sampled by the small particles during time we 
employ the following procedure: First we generate a sequence  of $N_\mathrm{c}$ configurations saved at equally spaced times $t_\mathrm{i}$ (with i=1$\ldots$ $N_\mathrm{c}$)
within a given time window $t_{N_\mathrm{c}}$.
The time interval $\Delta t_\mathrm{c}$ between two successive configurations, i.e. $\Delta t_\mathrm{c}= t_\mathrm{i+1}-t_\mathrm{i}$ 
is chosen in such a way that $\langle \Delta r^2(\Delta t_\mathrm{c})\rangle/\sigma_{\mathrm s}^2=0.5$.
Second, we overlap all $N_c$ configurations and perform a cluster size analysis according to the following
criteria:  (i) the  same particle at different times $t_i$ belong to the same cluster; (ii) if two particles overlap, they belong to the same cluster;
(iii) the size $s$ of a cluster is defined as the number of distinct particles belonging to the same cluster  (running from
one to the total number of small particles).
To improve statistics we average the cluster size distribution $n(s)$ over a set of at least 10 independent groups of $N_c$ configurations.
The average size of finite clusters is calculated as $L_\mathrm{c} = \sum s^2 n(s)/ \sum s n(s)$, excluding percolating clusters.

\begin{acknowledgments}
We thank Andrew Schofield (University of Edinburgh) for providing the PMMA particles, Vincent Martinez (University of Edinburgh), Wilson Poon (University of Edinburgh) and  Matthias Reufer (LSI instruments) for providing routines for DDM analysis, and Thomas Franosch (ITP Innsbr{\"u}ck) and Manuel A. Escobedo-S{\'a}nchez (University of D{\"u}sseldorf) for discussions. TS, SUE and ML acknowledge funding by the Deutsche Forschungsgemeinschaft (DFG) through the research unit FOR1394, project P2, and funding of the confocal microscope through grant INST 208/617-1 FUGG. 
EZ and CDM acknowledge support from MIUR through a Futuro in Ricerca grant FIRB ANISOFT (RBFR125H0M). 
EZ, CDM and FS acknowledge support from ERC-226207-PATCHYCOLLOIDS and ETN-COLLDENSE (H2020-MCSA-ITN-2014, Grant No. 642774).
\end{acknowledgments}

\section*{Author contributions}
$\S$ These authors contributed equally to this work.
TS, SUE and ML planned, performed, analysed and interpreted the experiments, EZ, PT, CDM and FS planned, ran and interpreted the simulations, TV obtained MCT predictions. All authors contributed to the interpretation and comparison of the data as well as the writing of the manuscript.

\section*{Additional Information}
Competing financial interests:
The authors declare no competing financial interests.


\newpage

\begin{figure}[h]
\centering
\includegraphics[scale=0.21]{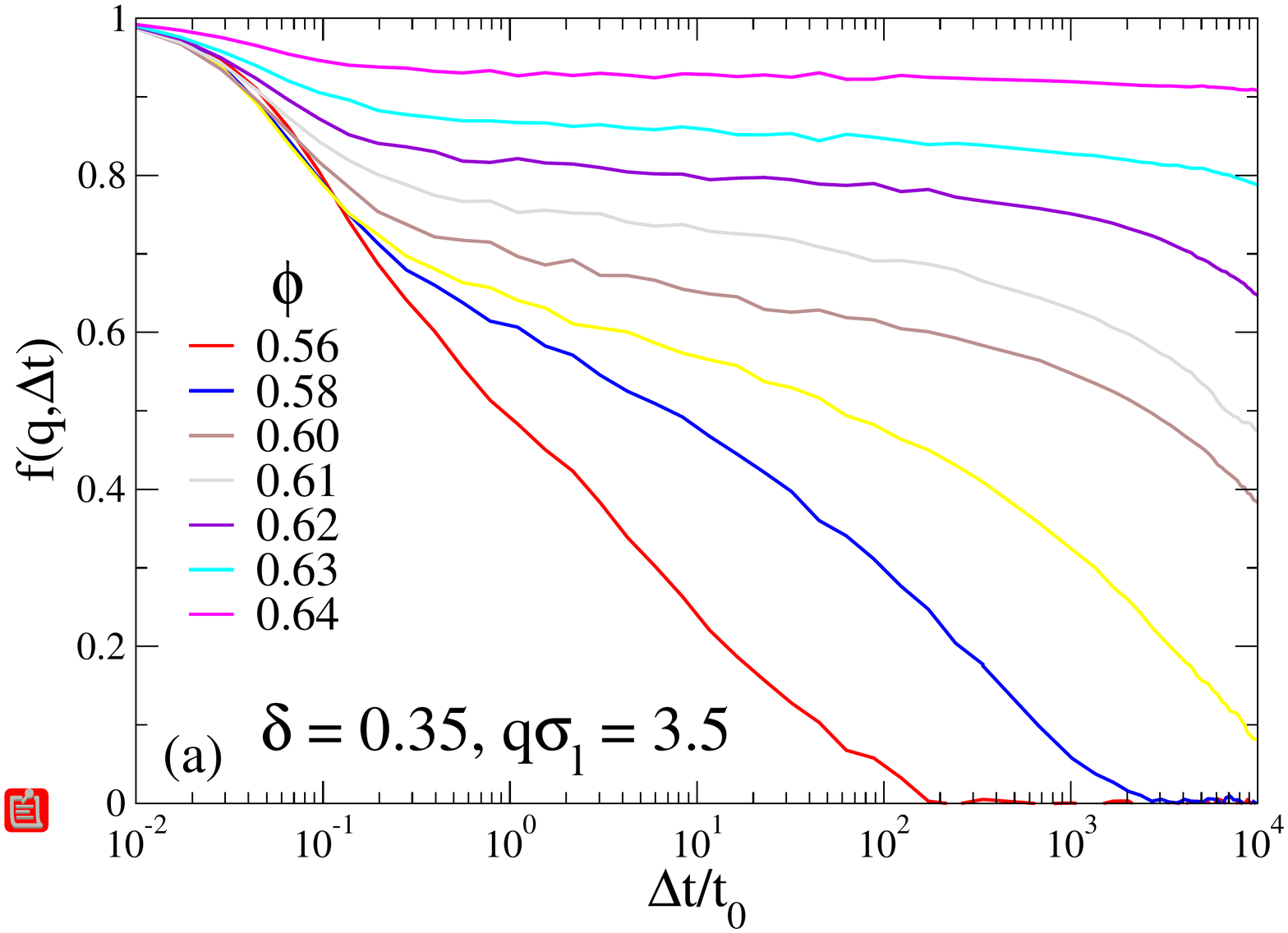}
\includegraphics[scale=0.21]{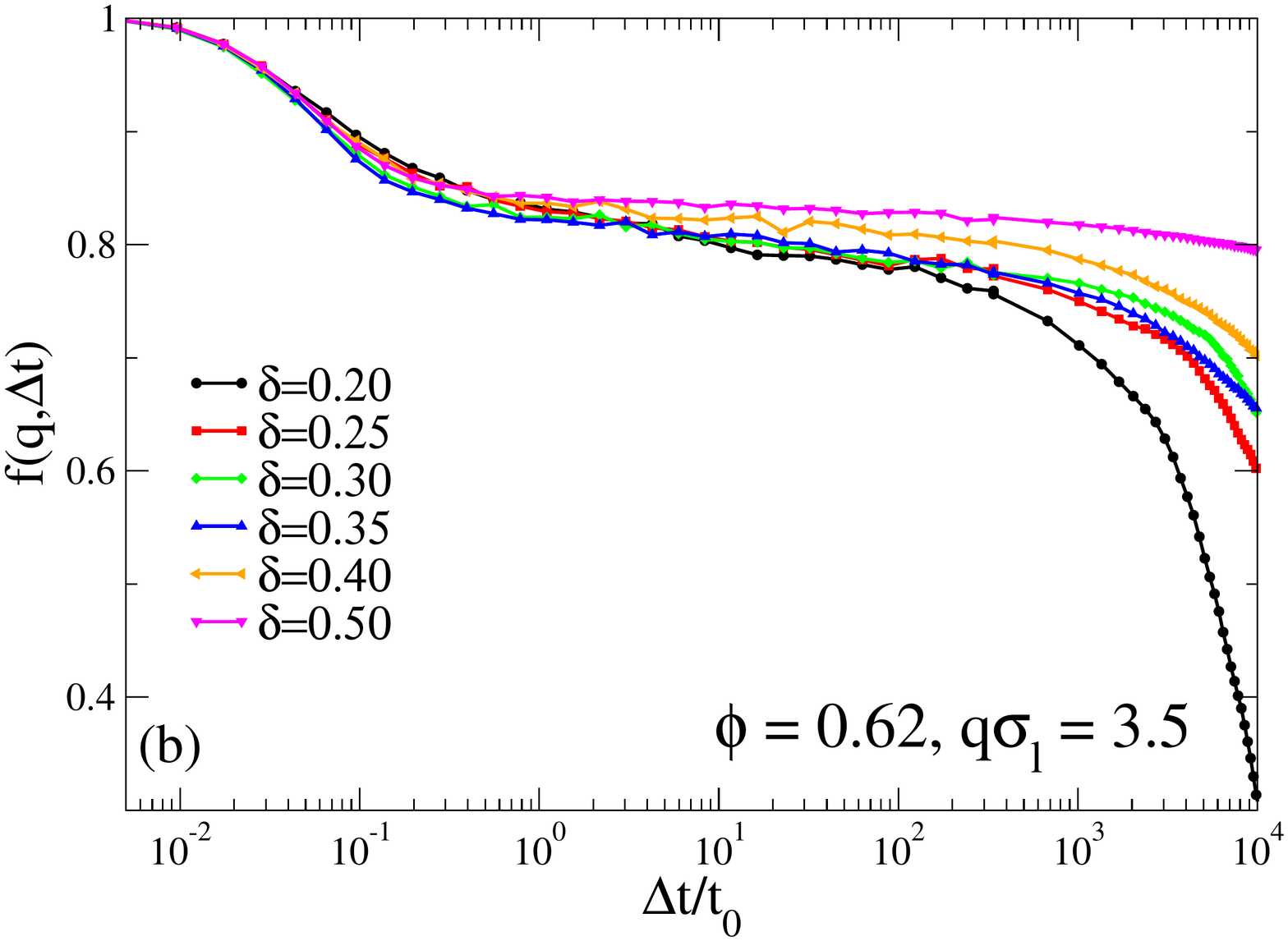}
\includegraphics[scale=0.21]{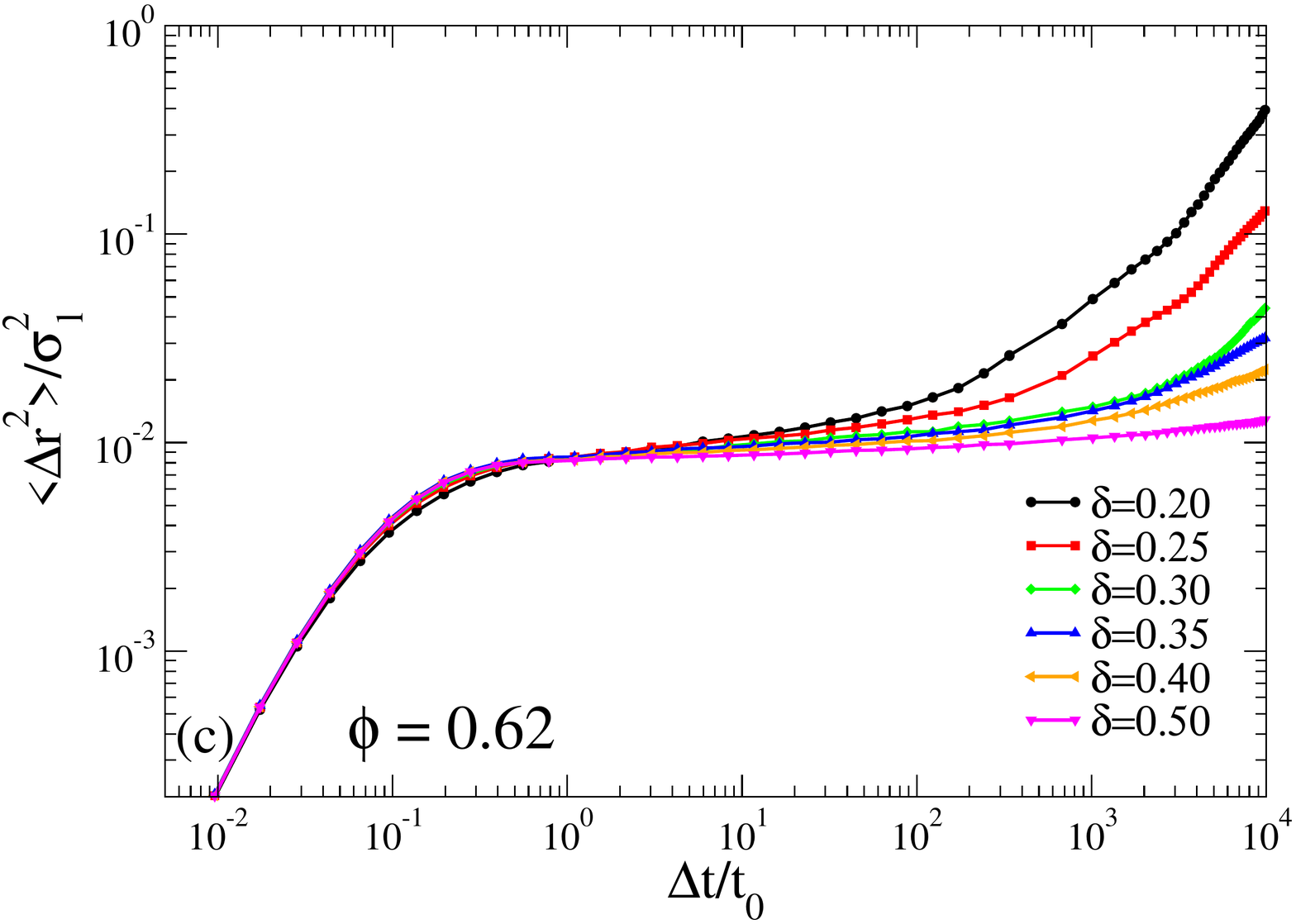}
\caption{{\bf Supplementary Figure 1. Large particle dynamics as obtained in simulations} Collective intermediate scattering functions  $f(q,\Delta t)$ at $q\sigma_\mathrm{l} \sim 3.5$ for (a) $\delta=0.35$, and different $\phi$ (as indicated) and (b) for $\phi=0.62$ and different $\delta$ (as indicated) (c) MSD for $\phi = 0.62$ and different $\delta$ (as indicated).
\label{fig:large}
}
\end{figure} 

\begin{figure}[h]
\centering
\includegraphics[scale=0.21]{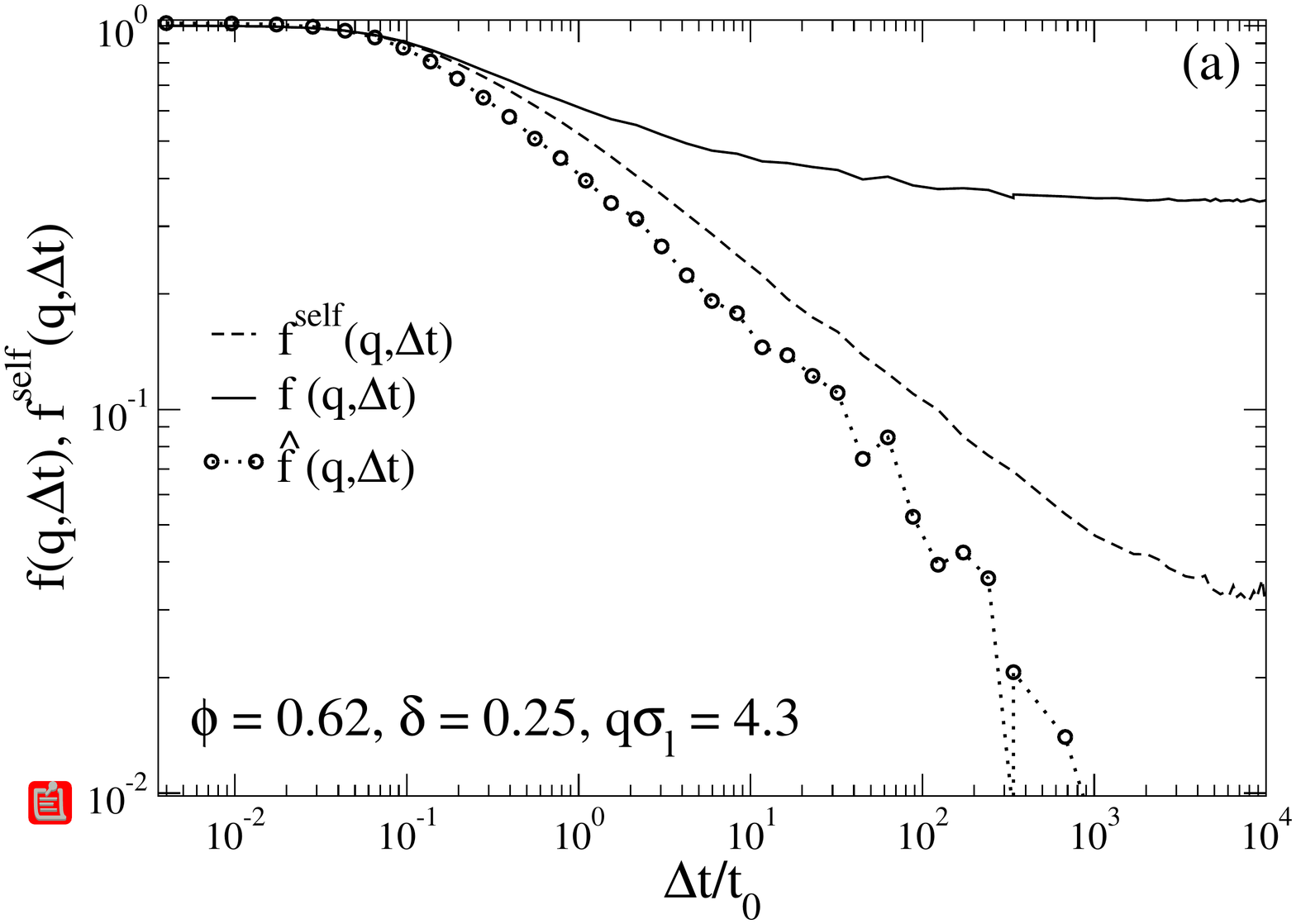}
\includegraphics[scale=0.21]{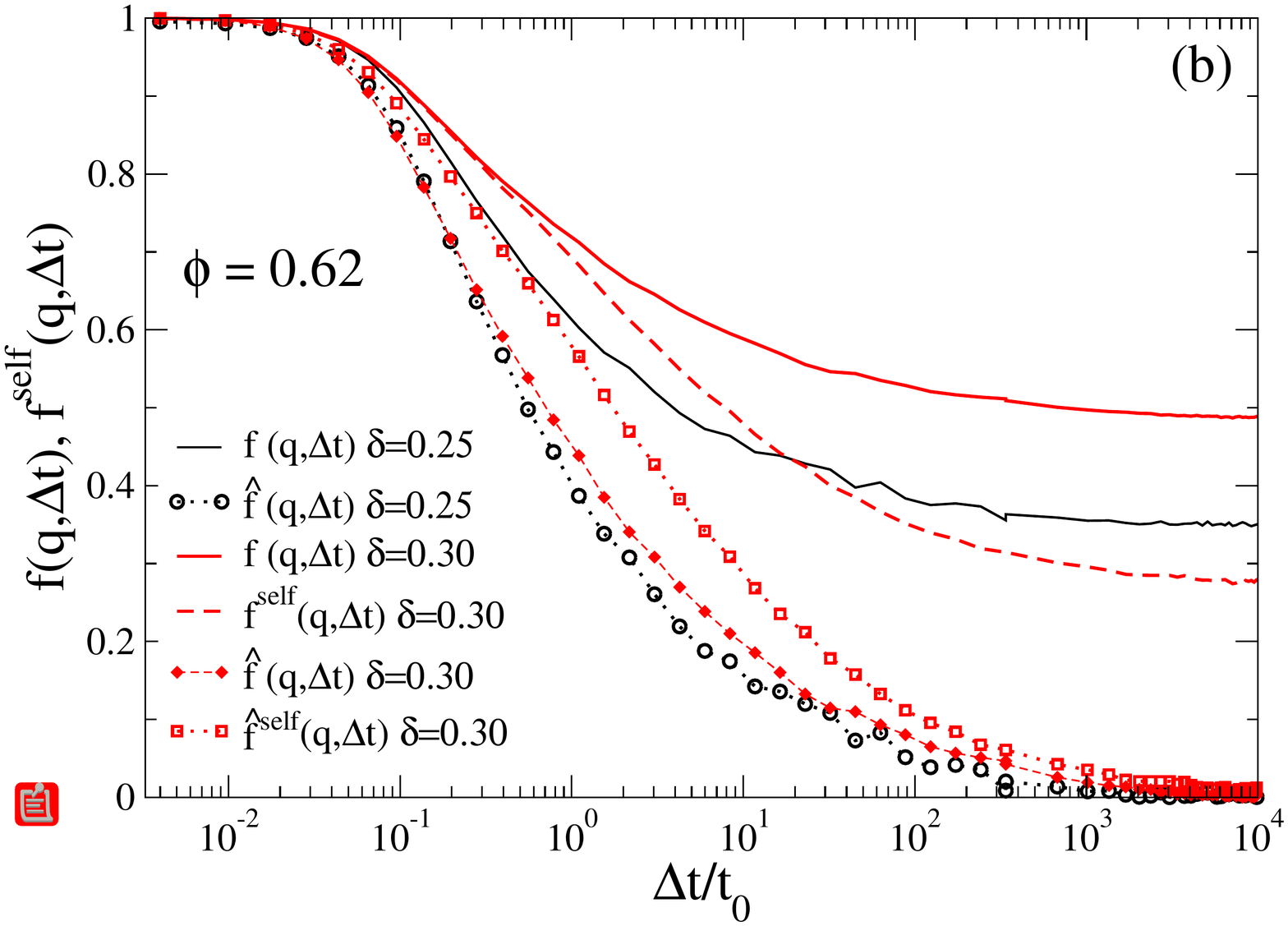}
\includegraphics[scale=0.21]{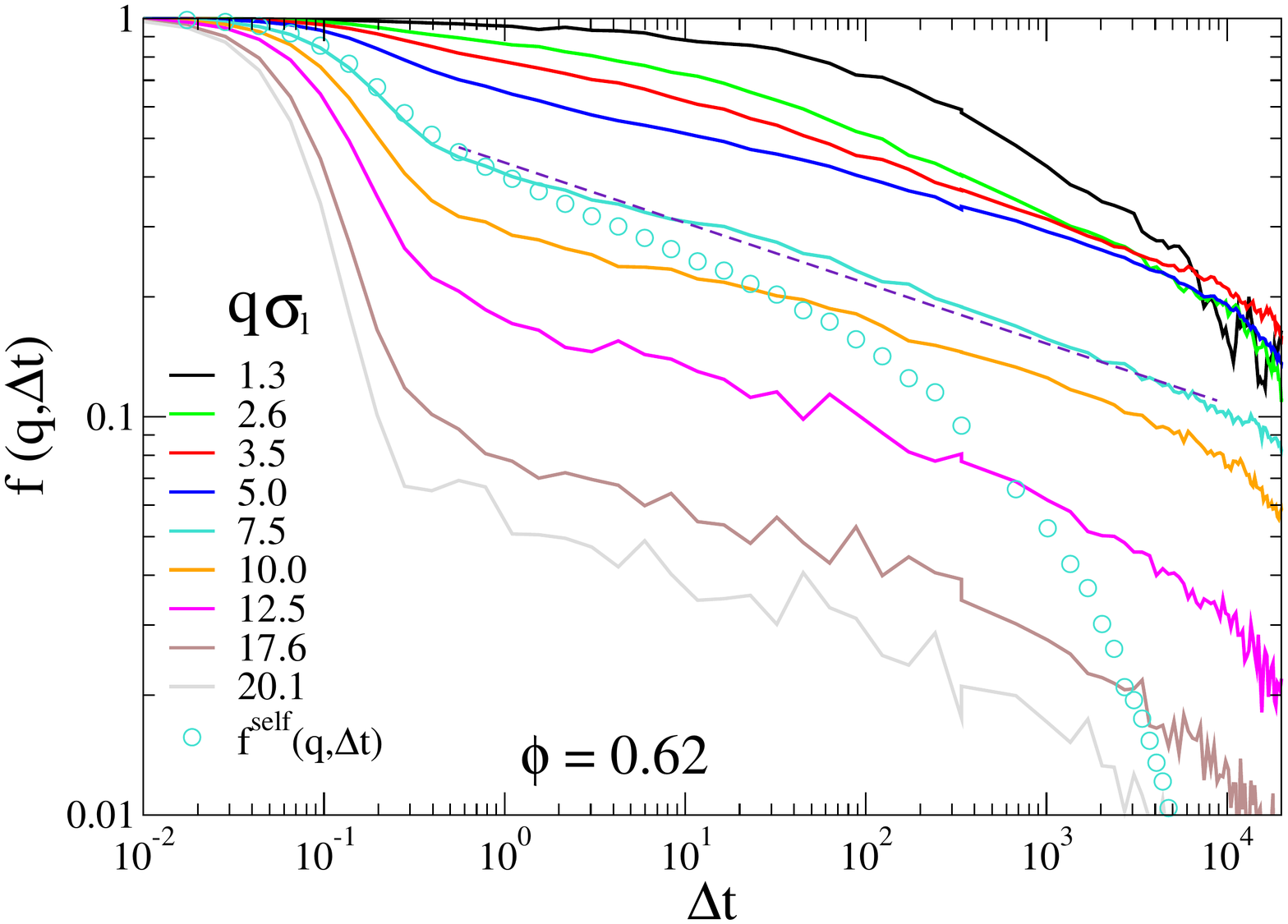}
\caption{{\bf Supplementary Figure 2. Small particle dynamics in an immobile matrix} (a) Collective (full curve), self (dashed curve) and scaled $\hat{f}(q,\Delta t)$ (see text, circles) intermediate scattering functions for $\phi=0.62$, $\delta=0.25$ and $q\sigma_\mathrm{l} \simeq 4.3$ in log-log plot; (b) collective (full curve), self (dashed curve) and scaled $\hat{f}(q,\Delta t)$ (symbols)
for $\phi=0.62$ and $\delta=0.25, q\sigma_\mathrm{l} \simeq 4.3$ (black),  $\delta=0.30, q\sigma_\mathrm{l} \simeq 3.7$ (red) in semi-log plot;
(c) Same data as in Fig.2h of the manuscript, but in log-log scale instead of semi-log scale: collective correlators for $\phi=0.62$ for various wavevectors. 
\label{fig:frozen}
}
\end{figure} 

\section*{Supplementary Notes}
{\bf Supplementary Note 1: Large particle dynamics}\\

The small particle dynamics display a dramatic change of behaviour at the critical size ratio, which can be associated with changes in the mechanism of arrest and the transition from caging at large $\delta$ to localisation at small $\delta$, related to the decoupling of the dynamics of the two species. On the other hand the arrest mechanism of the large particles, caging by other large particles, is not significantly affected by the presence of the small fraction, $x_\mathrm{s}= 0.01$, of small particles, irrespective of size ratio. As an example, Supplementary Fig.1(a) shows that large particles at $\delta_c=0.35$, where the small particles show anomalous dynamics, approach a standard glass transition upon increasing $\phi$, characterised by a typical two-step decay. Furthermore, Supplementary Fig.1(b) and (c) show that, upon changing $\delta$, the localisation length, i.e. the cage size, does not change significantly, as evident from the plateau height of both the MSDs ($\sim 0.1\sigma_{\mathrm l}^2$) and the density correlators. The cage though becomes more mobile with decreasing $\delta$, as shown by the faster dynamics at long times, indicating an interesting coupling between the increased mobility of the small and large particles with decreasing $\delta$. This coupling might be related to the fact that at small $\delta$ the small particles do not hinder the large particle movements due to their small size and large mobility. \\ 

{\bf Supplementary Note 2: Frozen vs. mobile matrix of large particles\\}

Here we want to compare simulations of a fully mobile binary mixture of hard spheres and one where the large particles are immobile. For the latter situation, {\it quantitatively} accurate results can only be obtained when one considers a large system size and also performs an average over several matrix realizations, as done in previous works \cite{coslovich,saito,franosch_frey}.
However, our aim is only to provide a {\it qualitative} comparison with the mobile case, for which our approach, based on a single realization for a system size of O($10^3)$ particles, is sufficient, as indicated by the fact that the MSD for the immobile matrix case reported in Fig.5(a) displays a qualitative behavior which is compatible with that of the Lorentz gas\cite{franosch_frey}.

Fig. 5(c) of the main article shows that the small particle self correlators for the immobile matrix case display (at intermediate time) a power-law behavior. It is to be noted that  these correlators, even below the critical size ratio $\sim 0.3$, display a long-time finite value, i.e. a residual non-ergodicity. 
Indeed, different from studies on the Lorentz gas \cite{franosch_softmatter},  we include among the intruders small particles 
trapped in finite size voids, i.e. not pertaining to the percolating cluster of voids, to make the analogy with the fully mobile case. The collective correlators, for the situation where the self ones show a power-law dependence on time, do not present the same behavior (Supplementary Fig.2(a)). Nevertheless, defining a scaled correlator $\hat{f}(q,\Delta t)=(f(q,\Delta t)-f(q,\infty))/(1-f(q,\infty))$, which allows us to remove the contribution of the frozen-in component to the correlation function \cite{Kertesz}, we see that a power-law behavior seems to emerge also for the collective correlators, even though our current numerical resolution is not good enough to determine this clearly.
However, the important point is that in semi-log plot (Supplementary Fig.2(b)) all correlators (self, collective and scaled) for frozen matrix conditions do not show a logarithmic decay
in any time window or wavevector. Finally, in Supplementary Fig.2(c) the correlators for the mobile matrix at the critical size ratio are reported in log-log plot showing that at $q\sigma_\mathrm{l} = 3.5$, where the anomalous logarithmic behavior is observed, a power-law decay cannot describe the data. It is interesting to note that at a larger value of $q\sigma_\mathrm{l} = 7.5$ the data might approach this behavior at long times, even though within a two-step decay. The power-law exponent of about 0.5 is also close to the Lorentz gas (0.527) and to MCT predictions. This suggests that at the smaller length scales probed at larger q values, the particles mainly see the local environment and localisation, while only at smaller q values the network structure of voids is explored and leads to anomalous behavior. Note though that the self correlators significantly deviate from power-law behavior. In summary, these results complement those provided in the manuscript and show that small particles moving in a frozen matrix behave very differently from those moving in a glassy but mobile matrix of  large particles.\\

\end{document}